\documentclass{elsart}
\usepackage{natbib}
\usepackage{epsfig}
\begin{document}
\begin{frontmatter}
\title{Dynamics of the decay $ \eta \rightarrow 3 \pi^0$}

\author[Indiana]{S.~Teige},
\author[RPI]{G.~S.~Adams},
\author[UND]{T.~Adams \thanksref{KSU}},
\author[UMD]{Z.~Bar-Yam},
\author[UND]{J.~M.~Bishop},
\author[MSU]{V.~A.~Bodyagin},
\author[NWU]{D.~S.~Brown},
\author[UND]{N.~M.~Cason},
\author[BNL]{S.~U.~Chung},
\author[UMD]{J.~P.~Cummings \thanksref{troy}},
\author[BNL]{K.~Danyo},
\author[MSU]{A.~I.~Demianov},
\author[IHEP]{S.~P.~Denisov},
\author[IHEP]{V.~Dorofeev},
\author[UMD]{J.~P.~Dowd},
\author[Indiana]{A.~R.~Dzierba},
\author[UMD]{P.~Eugenio \thanksref{cmu}},
\author[Indiana]{J.~Gunter},
\author[BNL]{R. Hackenburg},
\author[UND]{E.~I.~Ivanov},
\author[UMD]{M.~Hayek \thanksref{raf}},
\author[IHEP]{I.~A.~Kachaev},\
\author[UMD]{W.~Kern},
\author[UMD]{E.~King},
\author[MSU]{O.~L.~Kodolova},
\author[MSU]{V.~L.~Korotkikh},
\author[MSU]{M.~A.~Kostin},
\author[RPI]{J.~Kuhn},
\author[Indiana]{R.~Lindenbusch},
\author[IHEP]{V.~V.~Lipaev},
\author[UND]{J.~M.~LoSecco},
\author[UND]{J.~J.~Manak \thanksref{jlab}},
\author[RPI]{J.~Napolitano},
\author[RPI]{M.~Nozar},
\author[BNL]{C.~Olchanski},
\author[MSU]{A.~I.~Ostrovidov},
\author[NWU]{T.~K.~Pedlar},
\author[IHEP]{A.~V.~Popov},
\author[Indiana]{D.~R.~Rust},
\author[IHEP]{D.~I.~Ryabchikov},
\author[MSU]{L.~I.~Sarycheva},
\author[UND]{A.~H.~Sanjari},
\author[Indiana]{E.~Scott},
\author[UMD]{N.~Shenhav \thanksref{raf}},
\author[UND]{W.~D.~Shephard},
\author[Indiana]{M.~R.~Shepherd},
\author[MSU]{N.~B.~Sinev},
\author[NWU]{K.~K.~Seth},
\author[RPI]{J.~A.~Smith},
\author[Indiana]{P.~T.~Smith},
\author[UND]{D.~L.~Stienike},
\author[Indiana]{T.~Sulanke},
\author[Cyclotron]{A.~Szczepaniak},
\author[UND]{S.~A.~Taegar\thanksref{ASU}},
\author[UND]{D.~R.~Thompson},
\author[MSU]{I.~N.~Vardanyan},
\author[BNL]{D.~P.~Weygand \thanksref{jlab}},
\author[RPI]{D.~B.~White},
\author[BNL]{H.~J.~Willutzki},
\author[NWU]{J.~Wise},
\author[RPI]{M.~Witkowski},
\author[MSU]{A.~A.~Yershov},
and
\author[NWU]{D.~Zhao}

\address[Indiana]{Department of Physics, 
Indiana University,Bloomington IN 47405, USA}

\address[Cyclotron]{Nuclear Theory Center, 
Indiana University,Bloomington IN 47405, USA}

\address[BNL]{Department of Physics, 
Brookhaven National Laboratory,Upton, L.I., NY 11973}

\address[IHEP]{Institute for High Energy Physics,Protovino, Russian Federation}

\address[UMD]{Department of Physics, University of Massachusetts Dartmouth,
North Dartmouth MA 02747,USA}

\address[MSU]{Institute for Nuclear Physics, Moscow State University,Moscow, 
Russian Federation}

\address[NWU]{Department of Physics, Northwestern University,
Evanston IL 60208, USA}

\address[UND]{Department of Physics, University of Notre Dame, 
Notre Dame IN 46556, USA}

\address[RPI]{Department of Physics, 
Rensselaer Polytechnic Institute, Troy NY 12180,USA}


\thanks[KSU]{Now at Department of Physics, Kansas State University,
Manhattan KS 66506, USA}

\thanks[troy]{Now at Department of Physics,
Rensselaer Polytechnic Institute, Troy NY 12180, USA}

\thanks[cmu]{Now at Department of Physics, Carnegie Mellon University,
Pittsburg, PA 15213, USA}

\thanks[raf]{Permanent address: Rafael, Haifa, Israel}

\thanks[jlab]{Now at Thomas Jefferson National Accelerator
Facility, Newport News VA 23606, USA}

\thanks[ASU]{Now at Department of Physics,University of Arizona,
Tucson AZ 85721, USA}

\begin{abstract}
The parameter $ \alpha = -0.005 \pm 0.007 \;(stat) \pm 0.004 \;(syst)$ 
describing the density of events within the $ \eta \rightarrow 3\pi^0 $ 
Dalitz plot has been measured using data from Brookhaven
experiment E852. 
The result is compared to previous values
and theoretical predictions.
\end{abstract}
\end{frontmatter}

\section{Introduction}

Chiral perturbation theory offers a consistent 
description of low energy QCD particularly applicable
to meson decays. 
An outstanding difficulty is the
large discrepancy between the predicted 
and observed rate of the $ \eta \rightarrow 3\pi $
decay \cite{gasser}. 
We report here a measurement of the transition matrix 
element \mbox{$\mid \! M(\eta \rightarrow 3\pi^0) \! \mid^2$}.
The density of events on the Dalitz plot
has been predicted to be uniform 
by 
Gasser and Leutwyler \cite{gasser}
and by 
Di Vecchia \cite{vecchia}.
More recently a calculation of Kambor, Wiesendanger and Wyler 
\cite{kambor} found a density which decreases as the 
distance from the center of the plot increases.

The Dalitz plot density is specified by a single
parameter, conventionally denoted as $ \alpha$.
Kambor {\em{et al.}} show that both $ \alpha$
and the rate $ \Gamma (  \eta \rightarrow 3\pi  ) $
are sensitive to a parameter of the theory
( $ \overline{c} $ in their notation ).
The dependence of $ \alpha $ and $ \Gamma $ on $ \overline{c} $ 
is such that a more negative value of $ \alpha $
reduces the value of $ \Gamma $ thereby 
increasing the disagreement with experiment.
Two values of $ \alpha $ are determined by the authors of \cite{kambor},
-0.014 and -0.007, the latter containing higher
order corrections and leading to a prediction for
$ \Gamma (  \eta \rightarrow 3\pi  ) $ more
consistent with experiment.

To second order in the center of mass pion energy \cite{zemach} - \cite{cbar}
the Dalitz plot density for the decay 
$  \eta \rightarrow 3\pi^0 $  is parameterized as
\begin{equation}
\mid \! M \! \! \mid^2 \, \propto 1 + 2 \alpha z
\label{matdef}
\end{equation}
with
\begin{equation}
z = \frac{2}{3}\sum_{i=1}^{3} \left( 
\frac{3 E_i - m_{\eta}}{m_{\eta} - 3m_{\pi^0}} \right)^2 =
\left( \frac{\rho}{\rho_{max}} \right)^2
\label{zdef}
\end{equation}
where $ E_i $ is the energy of the $ i $'th pion 
in the $ \eta $ rest frame, $ \rho $
is the distance from the center of the Dalitz plot
and $ \rho_{max} $ the largest kinematically allowable
value of $ \rho $.

\begin{figure}[ht]
\begin{center}
\mbox{\epsfig{file=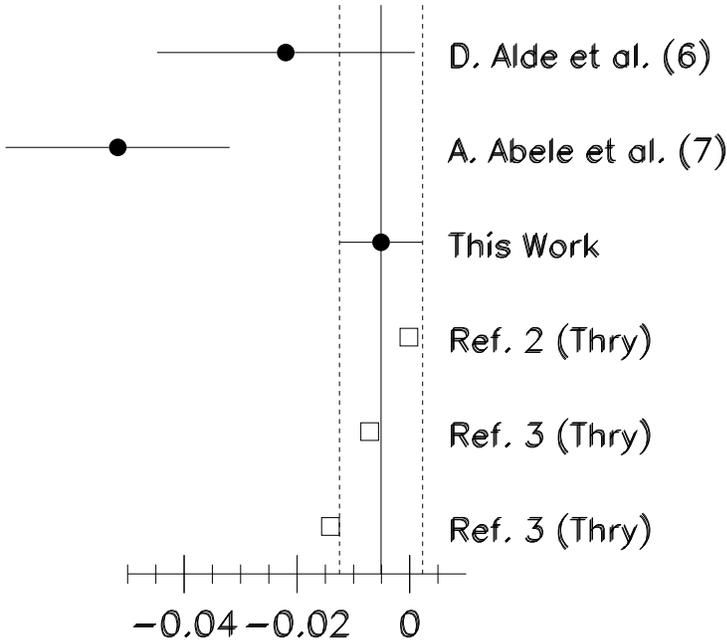,width=0.85\textwidth,height=0.85\textwidth}}
\end{center}
\vspace*{-0.8in}
\caption{
Previous measurements of $ \alpha $ and the result of this work
with a comparison to recent theoretical predictions. 
Not shown is the result of reference \cite{baglin}.
\label{systfig}}
\end{figure}

A uniformly populated Dalitz plot,
(as predicted by \cite{gasser} and \cite{vecchia}) 
gives $ \alpha =0 $.
$ \alpha $ has been measured previously by
Baglin {\em{et al.}} \cite{baglin} , 
Alde {\em{et al.}} \cite{alde} , 
and 
Abele {\em{et al.}} \cite{cbar}.
Recent results are shown in figure \ref{systfig}
along with the result of this work. 
Reference \cite{cbar} finds a value significantly different from zero 
and in probable disagreement
$ (2\sigma ) $ with theoretical expectations. 
The results reported in \cite{baglin} and \cite{alde}
are consistent with both zero and the non-zero expectation of \cite{kambor}.

In this work we report on a determination of $ \alpha $ 
from the decay $ \eta \rightarrow 3\pi^0 $ where the $ \eta $
was produced by 18.3 GeV/c negative pions incident on
liquid hydrogen.
The data were collected in 1995 by the Brookhaven
National Laboratory E852 collaboration.
The apparatus has been described previously \cite{brabson},\cite{teige}
and consisted of a large, segmented lead glass calorimeter (LGD), charged
particle tracking, veto and  trigger systems capable
of determining the charged particle multiplicity and the total
energy deposited in the lead glass.

The data used for this analysis were collected with the ``all-neutral''
trigger which required no charged particles, an energy deposition
greater than 12 GeV in the lead glass and no photons in the
downstream photon veto.

\begin{figure}[ht]
\begin{center}
\mbox{\epsfig{file=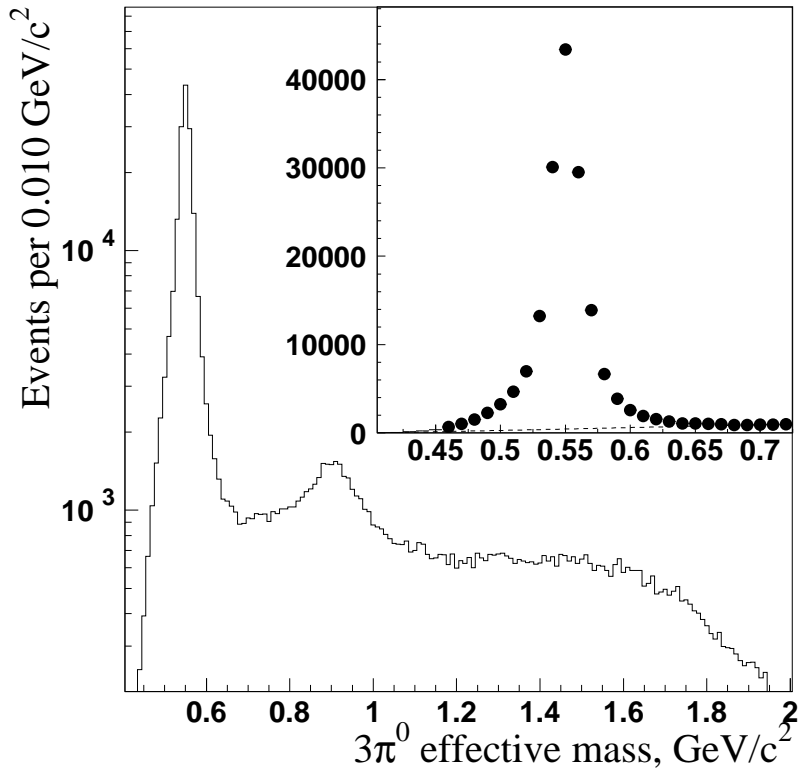,width=0.85\textwidth,height=0.85\textwidth}}
\end{center}
\vspace*{-0.2in}
\caption{The observed $ 3\pi^0 $ effective mass distribution.
The $ \eta $ signal dominates the data with
a secondary peak at $ \approx $ 0.9 ${\rm{GeV/c^2}}$ due to the 
decay $ {\rm{K}}^{*0} \rightarrow{\rm{K}}^{0}_{s} \pi;
{\rm{K}}^{0}_{s} \rightarrow  2\pi^0 $ visible. 
The inset shows the $ \eta $ region. There are 161,000 
events under the peak with an estimated signal to noise ratio of 
100 at the mass of the $ \eta $. The estimated background is
indicated by the dashed line.
\label{3pimass}}
\end{figure}

$ \eta \rightarrow 3\pi^0 $ decays were selected by requiring
exactly 6 reconstructed clusters in the LGD and a visible
total energy greater than 16.5 GeV. 
All combinations of assignments to the hypothesis $ 3\pi^0 $ were 
tested and if the best $ \chi^2 $ corresponded to a confidence level
greater than 1\% the event was considered a candidate. 
A preliminary fit using only the $ \pi^0 $ mass constraint was 
performed to allow evaluation of the $ 3\pi^0 $ effective mass. 
Figure \ref{3pimass} shows the resulting distribution.
Events with a mass less than 0.65 ${\rm{GeV/c^2}}$ (corresponding
to $ 3 \sigma $ from the $ \eta $ mass) were subjected to 
a full kinematic fit with the hypothesis  
$ \pi^- p \rightarrow \eta {\rm{n}} $; $ \eta \rightarrow 3\pi^0 $.
A confidence level (CL) cut selected decays for the final analysis.

It also was required that no photon have an energy 
less than 0.25 GeV and that the minimum separation between photons
be larger than 9 cm. 
The effect and purpose of these requirements is discussed below. 
The final resulting data set contained 87,500 events.

To determine $ \alpha $, eq. \ref{matdef} is fitted to
the observed distribution of $ z $ corrected for acceptance and phase 
space dependence.
The acceptance correction was based on Monte-Carlo simulation,
taking into account the apertures of the apparatus
and a large sample of GEANT generated electromagnetic
showers. The requirements on the minimum photon energy
and the minimum allowable photon separation mentioned above
removed any uncertainties in the Monte-Carlo simulation
associated with the transverse development of electromagnetic
showers and the behavior of low energy photons.
The separation requirement also selected events
where the nearest two electromagnetic showers were
well resolved. The phase space dependence of z
was removed by dividing the observed distribution by the
distribution due to a uniformly populated Dalitz plot.

\begin{figure}[ht]
\begin{center}
\mbox{\epsfig{file=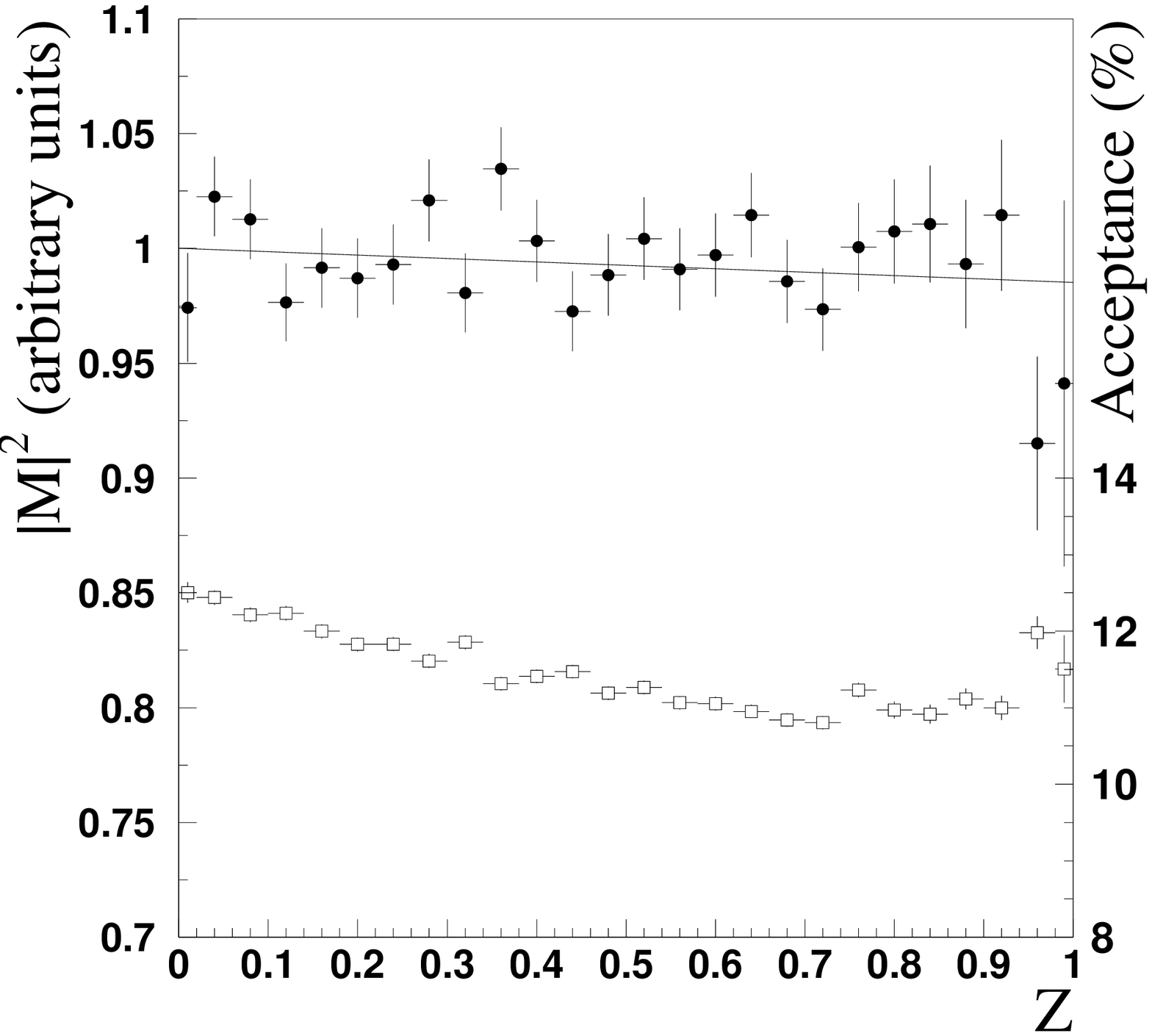,width=0.85\textwidth,height=0.85\textwidth}}
\end{center}
\vspace*{-0.2in}
\caption{Solid circles: z-distribution of $ \eta \rightarrow 3\pi^0 $ events
on the Dalitz plot. The solid line is the result of a fit
corresponding to eqn. \ref{matdef}. Open squares: the experimental
acceptance as a function of z.
\label{zdist}}
\end{figure}

The distribution resulting from removing
the phase space dependence and correcting for the
acceptance is proportional to $ M^2 $ and is shown in figure 
\ref{zdist} along with the result of the fit to eq. \ref{matdef}. 
The distribution has been numerically scaled to give a 
proportionality constant of one.
Our result is $ \alpha = -0.0047 \pm 0.0074 $, where the error 
is statistical only.

Systematic effects have been considered and the important
contributors identified.
Since $\alpha$ is a static property of the $ \eta $
a measurement of $\alpha$ cannot depend on
any data selection requirement, for example,
momentum-transfer, separation of photons in the detector, 
photon energy detection thresholds, confidence level of fits, etc.
It was observed that our measured $\alpha$ remained constant as a
function of the Confidence Level (CL) cut, down to about CL = 30\%,
below which $\alpha$ began to fall slightly.  This is consistent with an
increasing contribution from reconstructed events with one or more
wrong-assignments of a $\gamma$ to a $\pi^0$.  It was required
the an event have a confidence level greater than 30\% for this
analysis. Studies of the variation of $\alpha $ with the value
chosen for the CL cut yielded an estimate of $ \pm 0.003 $ for the
systematic contribution associated with this cut.

The analysis further required that no two photons have reconstructed
impact positions closer than 9 cm, a distance denoted by 
$ \Delta {\rm{r}} $.
Variation of the value of  $ \Delta {\rm{r}} $ used in the analysis
between 8.5 and 11 cm yielded our estimate of $ \pm 0.002 $
for the contribution to the systematic error due to this effect.

The effect of requiring the lowest energy photon to
have an energy larger than 0.25 GeV was investigated
by removing this requirement
(increasing the data set by 1,140 events) and by replacing it with
a requirement of 0.5 GeV
(decreasing the data set by 15,230 events). 
A contribution of $ \pm 0.001 $ to the systematic error was found.

For 20\% of the events in the data sample it was possible
to choose more than one assignment of photons to pions.
The analysis presented here chose the assignment with the
best confidence level. The effect of this choice was studied
by choosing the {\em{second}} best assignment (when possible)
and by using {\em{all}} assignments with an appropriate
weight to calculate $ \alpha $. No statistically significant
differences were observed.

Finally, the data set was divided into two statistically
independent halves by experimental run numbers and, separately,
by range of momentum transfers. In both cases, the values determined
for $ \alpha $ were not statistically significantly
different leading us to conclude these selections do
not contribute to our systematic error.
All effects considered combine to give our final
estimate of $ \pm 0.004 $ for the systematic error.

\section{Conclusion}

We have measured the $ \alpha $ parameter of the
transition matrix element $ M(\eta \rightarrow 3\pi^0) $
to be $ \alpha = -0.005 \pm 0.007 \;(stat) \pm 0.004 \;(syst) $.
Our result differs from one of the previous measurements \cite{cbar}
by slightly more than $ 2\sigma $ but is consistent with
several other \cite{alde}, \cite{baglin} measurements.
When compared to theoretical expectations, the value
of zero favored by Di Vecchia, Gasser and  Leutwyler
\cite{gasser} , \cite{vecchia},cannot be ruled out.
The result reported here is consistent
with the two values (-0.007 and -0.014) reported by 
Kambor, Wiesendanger and Wyler \cite{kambor}.

\end{document}